\documentclass[]{spie}  

\usepackage{CJK}
\usepackage{amsmath,amsfonts,amssymb}
\usepackage{graphicx}
\usepackage[colorlinks=true, allcolors=blue]{hyperref}

\title{Phase II of the Keck Planet Imager and Characterizer: system-level laboratory characterization and preliminary on-sky commissioning}

\newcommand{\caltech}{Department of Astronomy, California Institute of Technology, Pasadena, CA 91125, USA}

\newcommand{\ucsc}{Department of Astronomy \& Astrophysics, University of California, Santa Cruz, CA 95064, USA}
\newcommand{\keck}{W. M. Keck Observatory, 65-1120 Mamalahoa Hwy, Kamuela, HI, USA}
\newcommand{\ucla}{Department of Physics \& Astronomy, 430 Portola Plaza, University of California, Los Angeles, CA 90095, USA}
\newcommand{\jpl}{Jet Propulsion Laboratory, California Institute of Technology, 4800 Oak Grove Dr.,Pasadena, CA 91109, USA}
\newcommand{\ucsd}{Center for Astrophysics and Space Sciences, University of California, San Diego, La Jolla, CA 92093}

\newcommand{\berkeley}{Department of Astronomy, University of California at Berkeley, CA 94720, USA}

\author[a,$\star$]{Daniel~Echeverri}
\author[a]{Nemanja~Jovanovic}
\author[b]{Jacques-Robert~Delorme}
\author[a]{Yinzi~Xin}
\author[a]{Tobias~Schofield}
\author[c]{Luke~Finnerty}
\author[a,$\dagger$]{Jason~J.~Wang}
\author[a]{Jerry~Xuan}
\author[a,d]{Dimitri~Mawet}

\author[a]{Ashley~Baker}

\author[d]{Randall~Bartos}

\author[e]{Charlotte~Z.~Bond}

\author[f,$\ddagger$]{Marta L. Bryan}

\author[c]{Benjamin~Calvin}

\author[b]{Sylvain~Cetre}

\author[b]{Greg~Doppmann}

\author[c]{Michael~P.~Fitzgerald}

\author[a]{Jason~Fucik}

\author[a]{Katelyn Horstman}

\author[c]{Ronald~Lopez}

\author[g]{Emily~C.~Martin}

\author[d]{Stefan~Martin}

\author[d]{Bertrand~Mennesson}

\author[g]{Evan~Morris}

\author[a]{Reston~Nash}

\author[a]{Jacklyn~Pezzato}

\author[a]{Michael~Porter}

\author[b]{Sam~Ragland}

\author[a]{Mitsuko~Roberts}

\author[a,d]{Garreth~Ruane}

\author[a]{Jean-Baptiste~Ruffio}

\author[h]{Ben~Sappey}

\author[d]{Eugene~Serabyn}

\author[g]{Andrew~Skemer}

\author[i]{Taylor~Venenciano}

\author[d]{J.~Kent~Wallace}

\author[j]{Ji~Wang}

\author[b]{Peter~Wizinowich}

\affil[a]{\caltech}
\affil[b]{\keck}
\affil[c]{\ucla}
\affil[d]{\jpl}
\affil[e]{UK Astronomy Technology Centre, Royal Observatory, Edinburgh EH9 3HJ, United Kingdom}
\affil[f]{\berkeley}
\affil[g]{\ucsc}
\affil[h]{\ucsd}
\affil[i]{Physics and Astronomy Department, Pomona College, 333 N. College Way, Claremont, CA 91711, USA}
\affil[j]{Department of Astronomy, The Ohio State University, 100 W 18th Ave, Columbus, OH 43210 USA}

\authorinfo{$^\star$FINESST Fellow, Send correspondence to dechever@caltech.edu. $^\dagger$51 Pegasi b Fellow. $^\ddagger$NHFP Sagan Fellow.}

\graphicspath{{Figures/}}
\begin{document} 
\maketitle

\begin{abstract}
The Keck Planet Imager and Characterizer (KPIC) is a series of upgrades for the Keck II Adaptive Optics (AO) system and the NIRSPEC spectrograph to enable diffraction-limited, high-resolution ($R>30,000$) spectroscopy of exoplanets and low-mass companions in the K and L bands. Phase I consisted of single-mode fiber injection/extraction units (FIU/FEU) used in conjunction with an H-band pyramid wavefront sensor. Phase II, deployed and commissioned in 2022, adds a 1000-actuator deformable mirror, beam-shaping optics, a vortex coronagraph, and other upgrades to the FIU/FEU. The use of single-mode fibers provides a gain in stellar rejection, a substantial reduction in sky background, and an extremely stable line-spread function on the spectrograph.  

In this paper we present the results of extensive system-level laboratory testing and characterization showing the instrument’s Phase II throughput, stability, repeatability, and other key performance metrics prior to delivery and during installation at Keck. We also demonstrate the capabilities of the various observing modes enabled by the new system modules using internal test light sources. Finally, we show preliminary results of on-sky tests performed in the first few months of Phase II commissioning along with the next steps for the instrument.  

Once commissioning of Phase II is complete, KPIC will continue to characterize exoplanets at an unprecedented spectral resolution, thereby growing its already successful track record of 23 detected exoplanets and brown dwarfs from Phase I. Using the new vortex fiber nulling (VFN) mode, Phase II will also be able to search for exoplanets at small angular separations less than 45 milliarcseconds which conventional coronagraphs cannot reach. 
\end{abstract}


\section{INTRODUCTION}
\label{sec:intro}  
The Keck Planet Imager and Characterizer (KPIC) is an instrument designed to enable high spectral resolution characterisation of directly imaged exoplanets and brown dwarfs in the near-IR~\cite{Mawet2017_KPIC}. It uses a technique called high-dispersion coronagraphy (HDC)~\cite{Mawet2017_HDCII}. Deployed at Keck, the adaptive optics (AO) system first corrects for the atmosphere, and then light from the directly imageable companion is injected into a single-mode fiber (SMF) and routed to NIRSPEC, a high resolution spectrograph ($R>30,000$). KPIC is an interface between Keck AO and NIRSPEC. 

The deployment of KPIC was phased, with Phase I deploying in 2018. This included a new NIR pyramid wavefront sensor (PyWFS) to improve wavefront correction and allow for guiding on redder targets~\cite{Bond2020_PyWFS}. In addition, a simple injection system was deployed which consisted of a tracking camera for acquisition, a tip/tilt mirror guided by the camera to steer the target and align it onto the fiber tip, a bundle of SMFs to collect and transport the light, and an optical relay to reimage the beam onto the slit of NIRSPEC~\cite{Delorme2021_KPIC, Morris2020_KPICPhaseI}. The front-end dealing with coupling light into the fiber is referred to as the Fiber Injection Unit (FIU) while the back-end which reimages onto NIRSPEC is the Fiber Extraction Unit (FEU). This new HDC observing mode was commissioned between 2018 and 2020 at Keck and has successfully detected 23+ exoplanets and brown dwarfs to date, only a subset of which have been published thus far~\cite{Wang2021_KPICScience, Wang2022_KPICCORetrieval, Sappey2022_KPICHD206893, Wang2021_KPICPhaseI}. It has already enabled new science like the first spin measurements of the HR-8799 planets, robust abundance measurements, and the derivation of atmospheric properties from the combination of high and low resolution data. 

Phase II brings several upgrades to the system to improve planet throughput, minimize stellar leakage and enhance functionality. The first upgrade is a 1000-element deformable mirror (DM) from Boston Micromachines (BMC), which reduces fitting error and can boost the Strehl as well as enable speckle control at the location of the fiber. The second is a coronagraphic mechanism which supports a microdot apodizer mask to suppress stellar leakage~\cite{Calvin2021_KPICLab} and a vortex mask to enable vortex fiber nulling (VFN)~\cite{Ruane2018_VFN, Echeverri2019b_VFN}. Third is a set of Phase Induced Amplitude Apodization (PIAA) lenses~\cite{Guyon2003_PIAA}, which match the shape of the telescope beam to the fundamental mode of the fibers, thereby boosting coupling~\cite{Calvin2021_KPICLab, Jovanovic2017_SMFOnSky}. In addition to these key upgrades, the PyWFS and tracking camera pickoffs are now mechanisms allowing for several light sharing options, and the fiber port has been replaced with a stage that supports several lens/bundle combinations allowing KPIC to operate in other wavebands. 

Phase II was built in 2020 and 2021 and underwent extensive laboratory validation in late 2021/early 2022 before shipping to Keck for installation in February 2022. Here we provide an overview of the laboratory characterization and preliminary results from commissioning.

\section{PHASE II DESIGN}
\label{sec:design}
\begin{figure}[htbp]
\centering\includegraphics[width = \linewidth]{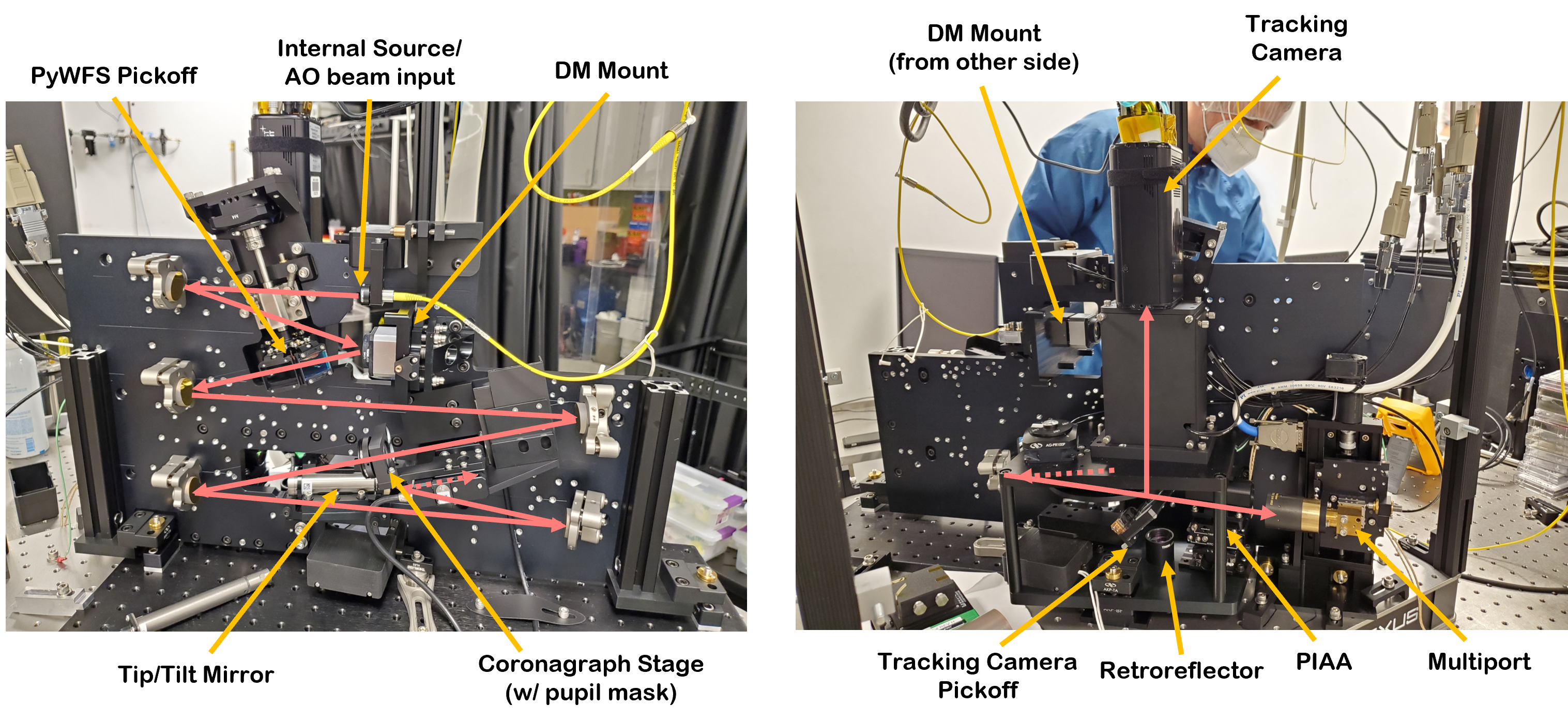}
\caption{A photo of the Phase II version of the KPIC instrument taken in the Caltech Exoplanet Technologies Laboratory before deployment to Keck in February 2022. (Left) Front side of the instrument. (Right) Rear side of the instrument.   \label{fig:instrument}}
\end{figure}

For a detailed description of the Phase II optical design we refer the reader to our previous work~\cite{Jovanovic2020_KPICPhaseII}. In this section, we provide a snapshot overview of the final as-built instrument. The instrument is shown in Fig.~\ref{fig:instrument}. The design philosophy is consistent with that of Phase I so that the Phase I plate could be easily replaced with the Phase II variant: namely that we used a similar optical layout (beam heights off the plate and bench), the same OAP prescriptions and relays, and the same vertical layout and kinematic features enabling a direct swap of the two plates at the observatory. Unlike the Phase I plate, the Phase II opto-mechanics were black-anodized to minimize stray light scattering and reflection.  

The left panel in Fig.~\ref{fig:instrument} shows the instrument as viewed from the front side. In the image, the input focal plane is occupied by the internal source fiber as  highlighted but this source can be retracted from the optical axis to allow the beam delivered by Keck AO to enter the system. When inserted, the source fiber allows the system to be internally calibrated independent of Keck AO, which was determined to be a valuable additional capability from Phase I. 

The injected beam is collimated by an OAP, and the pupil relayed onto the 1k DM. A flat mirror was temporarily mounted in a similar way to the DM so it could be rapidly swapped to simplify initial alignment. The beam then passes to the PyWFS pickoff which sends light in the wavefront control band to the Pyramid. This has been upgraded in Phase II to a rotating mechanism that offers 4 possible dichroics/mirrors to select which band the PyWFS uses for wavefront control. Switching wavefront control bands from the default H-band is necessary to observe with KPIC in that band and other short-wavelength bands. The transmitted beam is re-imaged via an OAP relay to a downstream pupil plane where the coronagraph mechanism is located. The mechanism consists of 3 options: a through-port for non-coronagraphic imaging, a microdot apodizer, and a vortex mask. The mechanism supports each of these by translating the given mask into, and out of, the beam and allowing for fine positioning to align it with the beam. Unlike other coronagraph implementations, KPIC chose to locate these optics in a pupil plane, simplifying the system and taking advantage of the fact pupil-plane VFN uses the same F/\# as for normal fiber injection into an SMF\cite{Ruane2019SPIE}. In a future upgrade, an atmospheric dispersion corrector (ADC) will be placed immediately after the coronagraph to correct for chromatic smearing and maximize broadband coupling into the fiber when observing a target away from zenith\cite{Wang2020_ADC}. This is still under development and will be deployed in 2023. Another OAP relay is used to re-image the pupil plane onto the tip/tilt mirror (TTM). The TTM is used to move the target around the fiber focal plane as needed to align it with the optical fiber of choice.  

Due to space constraints within the Keck AO bench, KPIC uses both sides of the vertical plate. The right panel in Fig.~\ref{fig:instrument} shows the rear side of the instrument with the TTM also serving to send light from one side to the other. The reflected beam is first incident on a flat mirror which sends the beam on its final trajectory towards the fiber focal plane. Before it reaches the fibers, the beam meets the tracking camera pickoff which has also been upgraded to a switchable, translating mechanism supporting up to three dichroics in Phase II. The light reflected from the selected pickoff dichroic is sent to the tracking camera which is used to visualize and determine the location of the target and drive the TTM to put the target on the fibers as desired. The tracking camera retains the ability to switch between focal plane viewing, used for tracking, and pupil plane viewing, used for aligning the coronagraphic masks. In addition, a Zernike mask is installed near the intermittent focal plane in the tracking camera optics box and enables Zernike wavefront sensing\cite{NDiaye2013_ZWFS, VanKooten2022_KPICZWFS}. Immediately before the tracking camera block is a filter wheel which allows the tracking wavelength to be changed as needed. 

The beam transmitted through the tracking camera pickoff is next incident on the PIAA lens assembly. This consists of two pre-aligned aspheric lenses in a tube which remap the pupil illumination from an obstructed flat top to a soft edged (quasi-Gaussian) beam that better matches the mode of the single-mode fiber. This optical system is on a stage allowing it to be inserted and retracted as needed. The final mechanism is called the ``multiport" which is a stage supporting up to three lenses and three corresponding bundles. This mechanism allows the observer to switch between various bundles which may be used for specific wavebands. KPIC Phase~I's default bundle was optimized for science in K and L band. An upcoming upgrade to Phase II will bring at least one more bundle optimized for science in y, J and H bands. 

The FEU was also upgraded in Phase II. For KPIC observations, NIRSPEC is moved into Keck AO and parked in front of the AO bench. In Phase I, KPIC required a daycrew member to connect the FEU-end of the fiber bundle to the FEU installed inside the calibration unit of NIRPSEC. This was prone to errors and increased the possibility that the fiber could be damaged either during the connection procedure or while it hung between the instruments during observation. To eliminate the need for a person to connect and disconnect the fiber before each run, and to reduce the likely hood of damaging the fibers, a new FEU was built inside Keck AO. It is located on the NIRSPAO plate. This is an optical relay that allows for the Keck AO beam to be sent directly to NIRSPEC. We installed the FEU such that we can inject the output of the bundles into the NIRSPAO optics and they project the image onto the slit of NIRSPEC. Specifically we use a mechanism very similar to the multiport to recollimate the output of the bundles, and a steering mirror located in the pupil plane to steer the beam across the slit of NIRSPEC. The steering mirror is mounted on a rotation stage which allows it to be removed and re-inserted into the beam. Additionally, there is a pickoff mirror that can be flipped into, and out of, the beam to send the light to a photodetector for daytime calibrations. This FEU upgrade has greatly simplified and sped up pre-observing preparations, improved the resulting calibration quality, and mitigated a key risk to the bundle recognized in Phase I.


\subsection{New Observing Modes}
\label{sec:new_modes}
With the upgrades and modes listed above, KPIC has expanded observing capabilities as described below. 
\begin{itemize}
    \item \textbf{Direct spectroscopy -} this is the classical HDC observing mode supported in Phase I where light from a known companion is directed onto, and injected into, the fibers in the bundle for spectral characterization. This mode offers a fiber for the science target, several fibers for the speckle field to get a contemporaneous spectrum of the host, and a background fiber. We usually bounce the target between two fibers during the observation to calibrate systematics. This mode works well for targets at $\gtrsim 5\lambda/D$. In Phase I, we successfully offset to a companion at 3.6 arcseconds ($80\lambda/D$). 
    \item \textbf{Apodized direct spectroscopy -} the microdot apodizer is designed to eliminate the majority of the static speckles from diffraction in the focal plane from 3-12 $\lambda/D$, thereby improving the detection of companions at these smaller separations. It does this at the expense of throughput, and with the increased losses, emissivity becomes an issue. As such the microdot apodizer was optimized for observations at shorter wavelengths (K band). To use the microdot apodizer, the mask is first inserted into the beam, and then the TTM steers the known companion onto the fiber of choice. 
    \item \textbf{Vortex fiber nulling (VFN) -} pushing for even smaller separations, this mode is optimized for the detection and characterization of targets between 0.5 and 2.0 $\lambda/D$ ($20-100$~mas in K-band). The vortex mask imprints a spiral phase ramp onto the beam which prevents on-axis light from coupling into the SMF. Off-axis objects within an annular "donut" profile, however, couple in and are transmitted to the spectrograph\cite{Ruane2018_VFN,Echeverri2019b_VFN}. Due to the azimuthal symmetry of the coupling region, new companions can be detected. VFN also enables spectral characterization of objects at spatial scales inaccessible to conventional coronagraphs. For VFN-mode observations, the host star is aligned with the core of the optical fiber once the vortex has been inserted in the pupil. Note that this mode does not provide information on the location of the companion around the host beyond the companions radial velocity.
    \item \textbf{PIAA -} this is not a standalone mode in its own right but acts as an enhancement to the other modes such that it can be combined with any of the previous 3 modes. The PIAA optics are used to boost coupling to the optical fiber. Although optimized across K and L band, the biggest gains come in L-band (boost from 65\% without PIAA to 86\% with PIAA). Given the PIAA uses a lossless apodization, it is optimal for L band observations. The PIAA does not improve speckle suppression. To use the PIAA, it is translated into the beam where it has been pre-aligned with one optical fiber. Then once the target is acquired, it is steered to that fiber. 
\end{itemize}








\section{LABORATORY VALIDATION}
\label{sec:lab_validation}
\begin{figure}[htbp]
\centering\includegraphics[width = \linewidth]{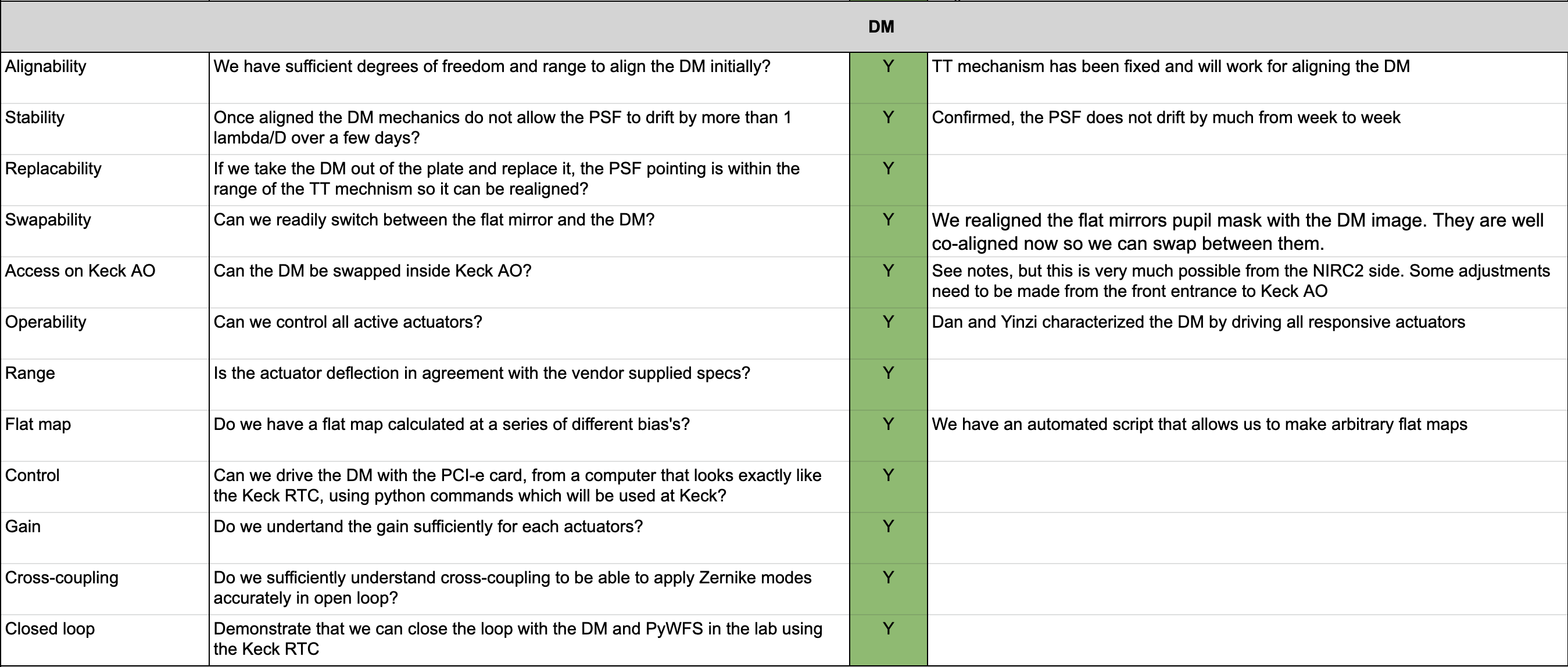}
\caption{The success criteria for the DM as an example of a single module's requirements and validation tests.   \label{fig:success}}
\end{figure}

Before deploying to Keck, we tested the Phase II system in the lab against a comprehensive list of success and pre-ship criteria and milestones developed to assess the performance of each sub-module, as well as the performance of the instrument as a whole. An example of the critera developed for the DM are shown in Fig.~\ref{fig:success} for reference. In this section, we report on some of the throughput and coupling efficiency measurements of the system. Note that this is just a small subset of the key results obtained in the laboratory. 

The throughput at 2 microns (a proxy for K band) from the input fiber focal plane to the fiber bundle focal plane was measured to be 72.6$\%$ without the PIAA and 69$\%$ with the PIAA. Note that this is the transmission of the optics up until the fiber bundle, not the coupling efficiency into the fiber; the losses with the PIAA in the beam path are due to the added optical surfaces but they are balanced out by the increased coupling efficiency from the mode matching that the PIAA provides. We can derive a theoretical throughput value from a bottom-up analysis where we account for the theoretical throughput of each optic. This yields an expected throughput of 5$\%$ higher than what was measured. Therefore our measured value is highly consistent with expectation. In addition, we measured the throughput from the input focal plane to the tracking camera with a 1550 nm laser to be 14$\%$, also consistent with expectation to within error. The choice of test wavelengths is consistent with the baseline direct spectroscopy observing mode where we observe in K and L and track in H band. 

\begin{figure}[htbp]
\centering\includegraphics[width = 0.8\linewidth]{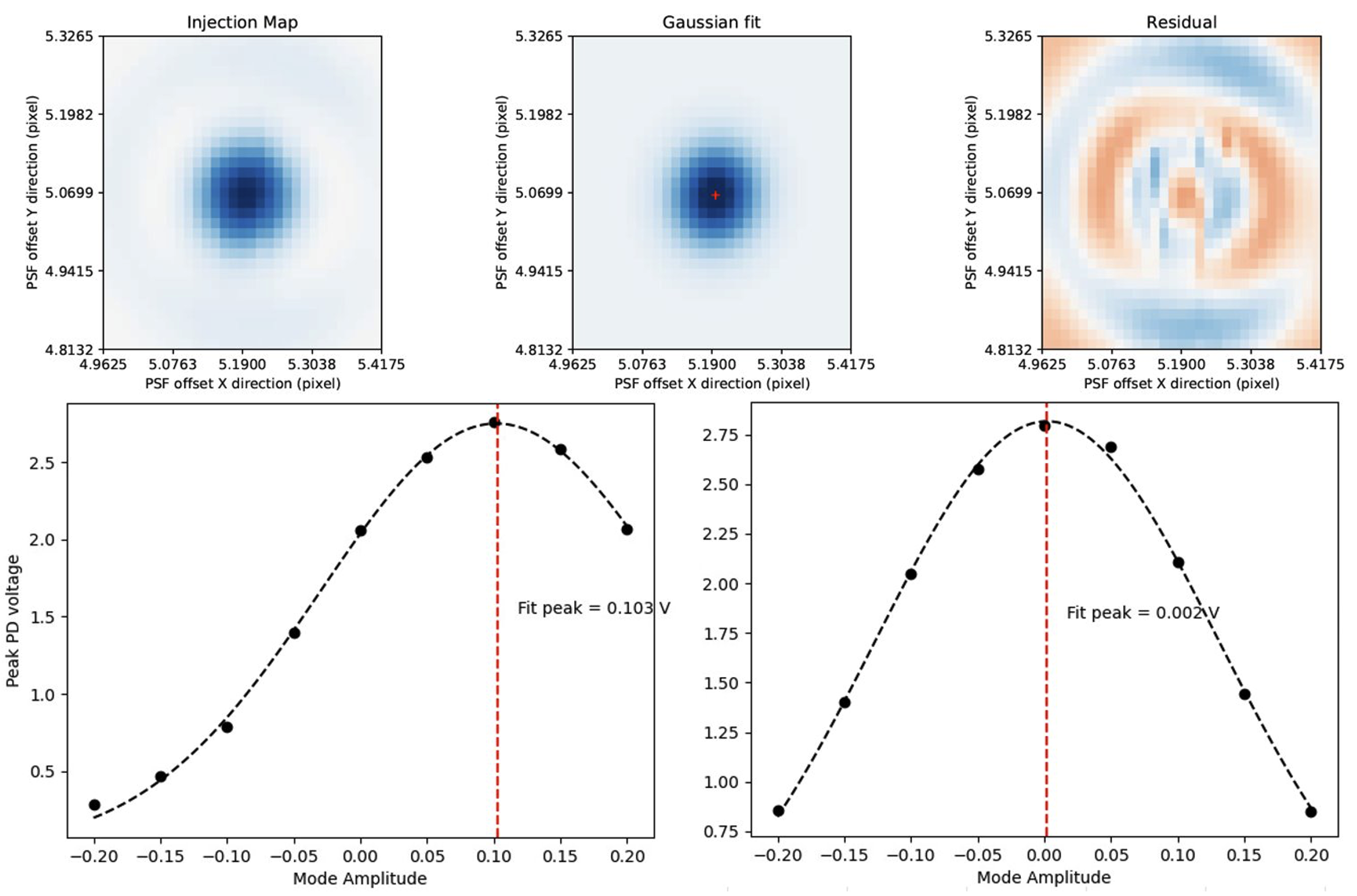}
\caption{(Top) Coupling maps made by scanning the TTM and recording the flux on the photodetector. (Left) Measured, (Center) Fit, (Right) Residual difference between measurement and fit. (Bottom) Flux measured on the photodetector as the defocus Zernike was scanned. (Left) Before correction, showing that some defous was present at the time. (Right) After applying the defocus correction there is a negligible amount of defocus left.    \label{fig:maps}}
\end{figure}

We then measured the coupling efficiency into a spare copy of the bundle used in KPIC during Phase I. Initially we placed a flat mirror in the location of the DM for this experiment, which allowed us to measure the coupling efficiency as limited by internal static aberrations. A photodetector was placed at the output of the bundle and the signal was sampled by the KPIC control computer. We conducted a Tip/Tilt (TT) scan with the mirror across the core of the fiber and measured the transmitted power. We fit this coupling map with a 2D Gaussian to find the peak and its location as seen in the top row of Fig.~\ref{fig:maps}. 
As the multiport provides the ability to adjust the distance between the focusing lens and the bundle, we made small adjustments to this gap and repeated the TT scan until we found the plane of best focus. The optimal coupling efficiency at 2 microns was measured to be 71-73$\%$. Given the pupil is unobstructed when using the internal source of KPIC, the theoretical limit is expected to be $\sim$80$\%$\cite{Shaklan1988}. After the flat mirror was replaced with the DM, we scanned the amplitude of the first 10 Zernike modes while monitoring the power on the photodetector. For each amplitude we applied for each Zernike, we conducted a full TT scan to compensate for any beam walk that may have occurred by applying the aberration. For each mode we plotted the peak flux from each coupling map as a function of the Zernike amplitude applied, and fit the the data with a Gaussian to find the optimum aberration amplitude to maximize coupling (see bottom row in Fig.~\ref{fig:maps}). We stepped through each Zernike mode in turn and improved the coupling efficiency up to 75-77$\%$. This demonstrated that we could achieve near perfect coupling efficiency in the laboratory with very little static wavefront correction required from the DM. 

Next we validated the stability of the injection, which is key to making high quality observations for a full night. 
\begin{figure}[htbp]
\centering\includegraphics[width = 0.8\linewidth]{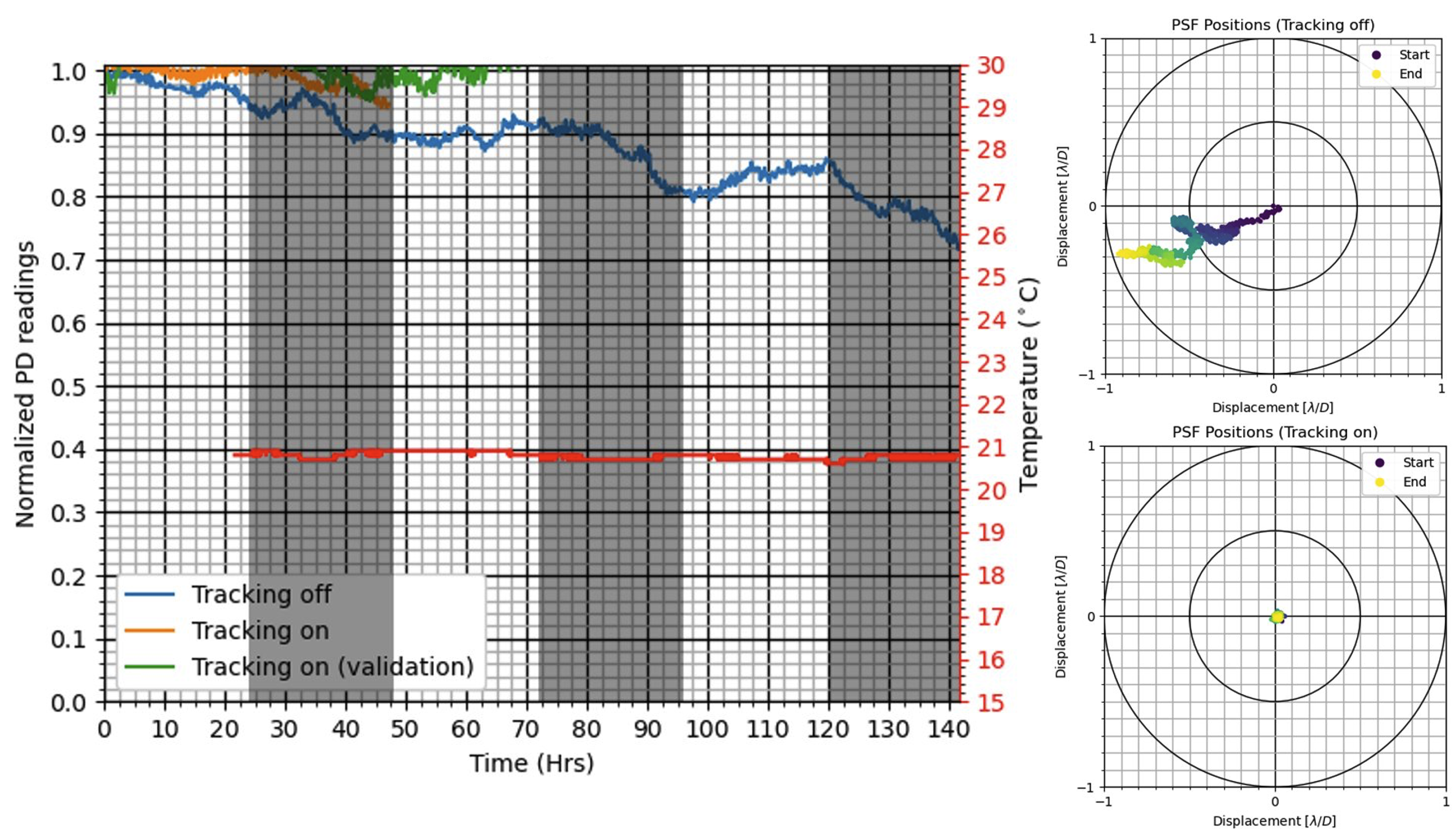}
\caption{System stability over the course of several days with and without tracking control loop active. (Left) Normalized photodetector readings through the bundle. Blue curve is with tracking control disabled while green and orange are repeat samples with tracking enabled. White and gray shading delimits 24-hour periods. (Right) PSF location on tracking camera relative to the start position with tracking loop off (Top) and with tracking loop on (Bottom).   \label{fig:stability}}
\end{figure}
With the coupling to the fiber optimized using the techniques described above, we tracked the flux evolution on the photodetector over a period of 6 days. The blue curve in the left panel of Fig.~\ref{fig:stability} shows the result. Over the course of the 6 days with no active control, meaning the system was allowed to drift and settle uncompensated, the power in the photodetector reduced by about 25$\%$. The location of the PSF is shown in the top right panel over that same period. It is clear that there is a drift of about $1\lambda/D$ in the location of the PSF at the tracking camera and this can be correlated with the drop in the flux. Nevertheless, given that an observing night is no longer than 12 hours, this demonstrates an extraordinary level of stability for the size of the fiber mode ($\sim$10 $\mu$m); over the first 12 hours of the test, there is a drop of less than 4$\%$. Further tests were conducted with KPIC's tracking loop which uses the tracking camera to determine the location of the beam and drives the TTM to actively maintain the alignment of the beam with a given pixel on the detector. The resulting PSF stability on the tracking camera with tracking contorl on is shown in the bottom right panel indicating no deviation in the beam. The flux during this test, which was repeated twice, is shown in the green and orange traces in the left panel and shows that there was no net drift once the tracking loop was closed using the tracking camera. This indicates that there is very little non-common path drift between the tracking camera and the fiber bundle, and hence we can calibrate the two before a science run and drive the system to maximal coupling for an extended period thereafter. 


\section{EARLY COMMISSIONING RESULTS}
\label{sec:early_commissioning}
The Phase II version of the FIU along with the new FEU were deployed to the summit of Keck on February 14\textsuperscript{th}, 2022. The installation required that KPIC Phase I and the PyWFS be removed entirely from the bench and placed on a breadboard in the AO room. With the instrument set up like this, we could swap the Phase II plate for the Phase I plate. In addition, we had to upgrade many of the Phase I electronics, install the new electronics in the electronics vault room, and route the cables to the AO room. All the new devices brought in Phase II had to be set so they could be controlled via terminal server at Keck as well as through Keck keywords. There were other upgrades and repairs also carried out on the PyWFS as well as NIRSPEC. The bulk of the summit work was performed in 3 weeks by a team of 4, with residual items mostly solved over the next few months. 

Since the install was completed in early March, the project has received 4 commissioning nights which are covered in this proceedings. There was a full week of science verification observations carried out simultaneous to the SPIE conference so some preliminary results from those nights were shown during the conference talk but, aside from two key highlights, the new data from that run of nights is not handled here; they will be covered in a later publication. As such, unless explicitly stated otherwise below, the results in this paper only deal with observations made in those initial 4 commissioning nights.

The observing conditions were not favorable for the majority of the commissioning time, including a half night with overhead clouds, one full night with $>2.5$ arcsec seeing and another half night where we were dominated by low wind effect. Despite this, the team has re-validated the Phase I capabilities and some of the Phase II capabilities. Due to some on-going issues with the PyWFS, all work thus far was carried out with the Shack-Hartman wavefront sensor and the Xinetics DM. Here we limit the summary to the recommissioning of Phase I capabilities, critical to getting the instrument back on-sky for science observations.  

\begin{figure}[t!]
\centering\includegraphics[width = \linewidth]{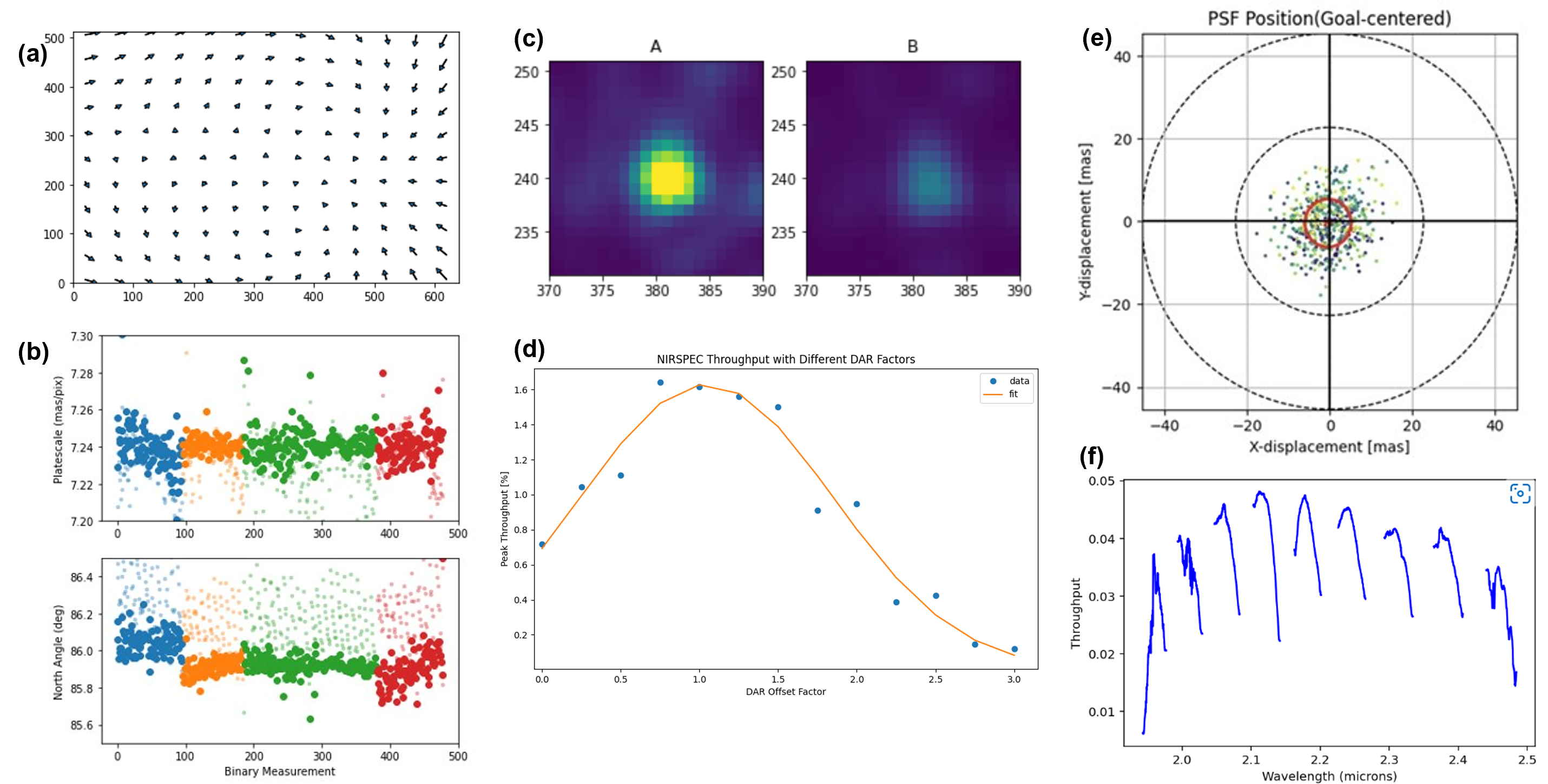}
\caption{(a) Distortion solution for the tracking camera. (b) Plate scale and North angle solutions. (c) Offsetting test results. The left plot shows the primary and the right plot the secondary. (d) The throughput to NIRPSEC as a function of the differential atmospheric refraction correction. (e) Centroid plot showing the tracking loop working on-sky. The red circle marks the standard deviation of all the centroids. (f) End-to-end throughput, including atmosphere, on NIRSPEC across k-band - this throughput is from the latest observing run (July 18\textsuperscript{th}, 2022).  \label{fig:commissioning}}
\end{figure}

To be able to acquire a target, the first step was to calibrate the North angle, plate scale and distortion solution for the tracking camera. This was done by imaging a set of binaries with the tracking camera as well as NIRC2. NIRC2 images were used to determine the separation and orientation of the binary system. Then for each binary, we used the TTM to move the system across the tracking camera in a 10$\times$10 grid. The resulting distortion solution is shown in panel (a) of Fig.~\ref{fig:commissioning}, while panel (b) shows the plate scale and North angle. The preliminary plate scale was determined to be 7.24 mas/pixel and the North angle 85.93 degrees. Given the importance of having accurate values for these metrics, the measurement will be repeated whenever possible. 

The next step in coupling light into the bundle was to demonstrate that the Phase I tracking loop can be used to acquire the target and steer it to the location of the fiber. The Phase I tracking loop captures 10 images on the tracking camera, averages them, fits a 2D Gaussian to the PSF to find its position, and then drives the TTM to steer the target to the desired pixel on the camera which has been identified to correspond with the desired fiber in the bundle. By using a 10-frame average, the loop runs as a drift-compensation loop at this stage. A plot of the distribution of centroid locations during a tracking sequence in H-band can be seen in panel (e). This shows the beam is well aligned with the required pixel throughout the test validating the Phase I tracking loop is operational in Phase II. 

To observe a faint companion, it is critical that the offsetting in the tracking loop, which is used to place the companion on the fiber, is done correctly. This is where the distortion map, plate scale, and North angle determined earlier play a key role. To validate the procedure, including the values above, we observed a series of low contrast binaries on NIRC2, which has a very well calibrated focal plane, solved for the PA and separation of the companion, and then input this information into our control loop. We first aligned the primary on the pixel/fiber of choice (left image in panel (c)) and then applied the offset to align the secondary with that same pixel/fiber (right plot of panel (c)). When working correctly, this procedure should result in the primary and secondary landing on the same pixels in Fig.~\ref{fig:commissioning}(c). For both sets, we used the tracking loop to keep the target aligned and collected data with the tracking camera to assess how close to the desired position the target landed. We have only conducted this on two targets thus far, and achieved a residual offset of 5 and 10 mas respectively, when offsetting to a target at 1.3 arcsec. We need to collect more data to determine if this is a random or systematic error. Nonetheless, the current levels of error are sufficiently small to proceed to science operations. 

The final key item to re-validate before commencing science operations was to check that the offsetting applied to compensate for differential atmospheric refraction (DAR) was correct. Since Keck does not have an ADC, when viewing away from Zenith the PSF is chromatically dispersed in the focal plane. The ADC we plan to deploy in 2023 will correct this. In the meantime however, the PSF seen by the tracking camera in H-band will be offset with respect to the optical fiber which operates at K and L band. To correct for this, we use a model of the DAR between these two bands and apply an offset with our TTM along the elevation axis to ensure that the center of K-band is aligned with optical fiber during an observing sequence. Panel (d) of Fig.~\ref{fig:success} shows the throughput to NIRSPEC (i.e. flux) as a function of the DAR offset factor. Assuming our model and calculations are correct, a DAR factor of 1 is the optimum factor for any observation. We applied offsets to this and collected data with NIRSPEC to confirm that we get a peak at a DAR offset factor of 1 and indeed we do, validating that offsetting to compensate for DAR is working well. 

In order to demonstrate that the Phase I capabilities are fully recovered in Phase II, we had to meet two milestones: (1) confirm that we have equal, if not better, throughput to NIRSPEC and (2) detect a large-separation companion to show that all offsetting procedures are operational. On the best nights with the Phase I system, we would obtain a peak end-to-end throughput, including the atmosphere and all optics through to the NIRSPEC detector, of 3\%\cite{Delorme2021_KPIC}. During the four Phase II commissioning nights, we demonstrated that we can now regularly hit 3\% throughput despite average observing conditions. Furthermore, during the more recent round of observing nights not generally covered in this proceedings, we demonstrated a peak throughput of 5\% multiple nights in a row. Figure.~\ref{fig:commissioning}(f) shows the end-to-end throughput as measured on-sky on July 18\textsuperscript{th}, 2022. This confirms point (1) by showing that we have not only recovered but in fact have close to doubled the throughput of the system despite not using the PyWFS or new high-order DM for active wavefront control. We believe most of this improvement is due to upgrades to some of the optics to improve throughput specifically as well as improved calibration procedures leveraging the new FEU to maximize coupling before going on-sky. For point (2), we recently observed (July 18\textsuperscript{th}, 2022 as well), and detected, HIP~79098B which is a system with a 20~Jupiter-Mass companion at 2.4 arcseconds. With this detection, we are confident that our offsetting procedures are working as well, if not better, than they did in Phase I.

As such, this confirms that KPIC has regained, and improved, its Phase I capabilities and is now ready for science while further commissioning of Phase II modules is conducted. 


\subsection{VFN Mode Commissioning}
\label{sec:VFN_comm}
One new Phase II mode which has had some on-sky commissioning time with notable results so far is the VFN mode. The goal of this mode is to observe targets at small separations between $0.5$ and $2.0\lambda/D$. With its symmetric coupling region, VFN can not only spectrally characterize known companions but can also detect new companions. During testing before deploying to Keck, we demonstrated monochromatic nulls of $2.1\times10^{-3}$ with simultaneous peak coupling efficiencies of 8.4\% at $1.4\lambda/D$ in 2 micron laser light in the lab. Once at Keck where we can use NIRSPEC to get a spectrally-resolved null, we demonstrated nulls of $<10^{-3}$ in parts of K-band during daytime testing. These tests are ongoing to determine the off-sky limits of the KPIC VFN mode but they were sufficient for us to start on-sky commissioning tasks.

\begin{figure}[htbp]
\centering\includegraphics[width = \linewidth]{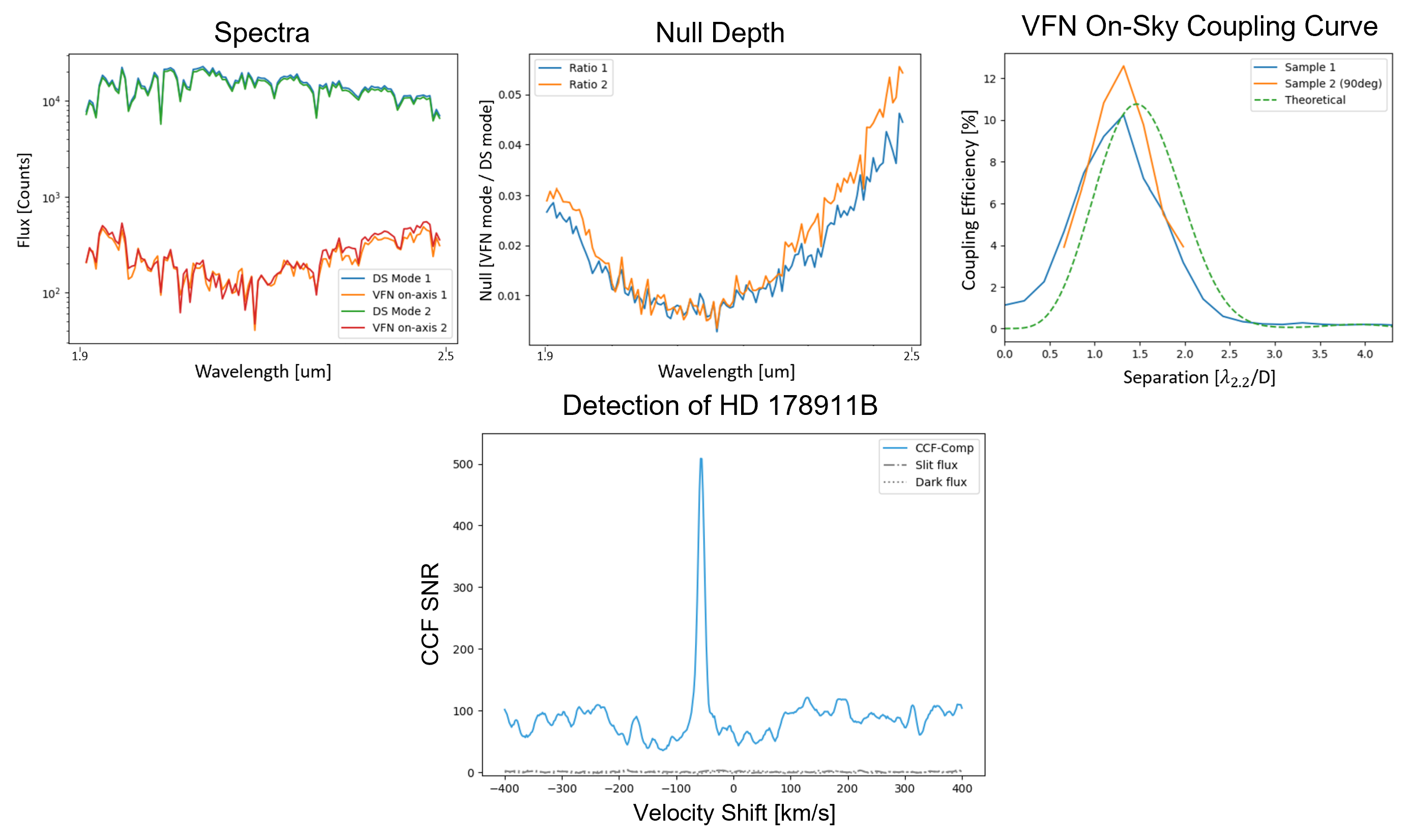}
\caption{Preliminary results of on-sky commissioning for the KPIC VFN mode. (Top left) Two samples each of the stellar spectrum in direct spectroscopy (DS) and VFN mode. (Top center) Ratio of VFN to DS mode flux showing a clear nulling of $\sim10^{-2}$. (Top right) off-axis coupling efficiency. (Bottom) Strong detection of a known low-contrast stellar binary at 50~mas with VFN. \label{fig:VFN}}
\end{figure}

The top row of Fig.~\ref{fig:VFN} shows the results of the first on-sky commissioning night with KPIC VFN. The top left panel shows the downsampled stellar spectrum across K band when the star is aligned with the fiber without the vortex mask (blue and green) as well as with the vortex mask (orange and red). This effectively compares the amount starlight coupling into the fiber in the direct spectroscopy (DS) mode versus the VFN mode. There is a clear reduction in starlight by 2 orders of magnitude as well as a noticeably deeper rejection towards the center of the band, reflecting the fact that the vortex works better closer to 2.2 microns. The top center panel shows the ratio of these two spectra which is an indirect measure of the null since it does not account for losses in the DS mode itself. Nevertheless, this null-like ratio is clearly on the order of $10^{-2}$ on-sky and has the expected parabolic shape. We also measured the off-axis coupling efficiency on-sky by scanning the star over the fiber in a line at two different angles. This simulates a companion at a different separation and PA corresponding to the offset applied to the star with respect to the fiber. The blue and orange curves are the measured values while the dashed green curve is the theoretical performance; there is a clear agreement between the two with an average peak coupling efficiency of about 10\% as expected. 

Having demonstrated that the VFN can indeed null on-sky and achieves the expected off-axis coupling, thereby verifying the concept on-sky for the first time ever, we moved to demonstrate a low-contrast detection. We chose HD~178911 which is a known stellar binary with a G1V primary and K1V companion\cite{Farrington2014_HD178911}. At the time of observation, the companion was at a separation of 50~mas ($\sim1.1\lambda/D$). From the spectral types, we can predict a V band flux ratio of about 1/7. The night of observation, conditions were poor such that the PSF was unstable and highly variable. However, we still achieved a very strong detection of the companion with a cross-correlation function signal to noise ratio (CCF SNR) of about 500 as shown in Fig.~\ref{fig:VFN}. A preliminary spectral analysis measured a relative radial velocity (RV) of 14~km/s which is in very close agreement with the expected 13.7~km/s at the time of observation from the accepted orbital solution. We also recovered an effective temperature and log(g) consistent with values predicted from the spectral type and mass of the objects but need to push on this analysis further before final values can be determined. 

With these null and coupling efficiency measurements as well as the detection of a low-contrast companion, we can confidently say that VFN is working on-sky and that our analysis procedure works well. We just need more on-sky commissioning and science verification time to push the VFN mode further.

\section{NEXT STEPS}
\label{sec:next_steps}

With the Phase I capabilities revalidated and the new Phase II modes progressing nicely toward on-sky verification, there still remain other short term tasks that the we are working on. The Phase I tracking control loop has proven to work very well as a drift-control system able to center targets on the fiber well enough to achieve consistent, high throughput measurements. However, the residual tip/tilt jitter of this control algorithm is on the order of 7-8~mas RMS such that there is still some room for improvement in the direct spectroscopy coupling efficiency and, more significantly, the VFN null will likely be limited by jitter very soon. As such, we are upgrading the control loop to use a new PSF finding algorithm which should be much faster and we are overhauling the code to minimize runtime and thus operate much faster, thereby specifically targeting higher frequencies and hence jitter.  Preliminary development and testing so far has shown a factor $50\times$ improvement in runtime with the new algorithm and we are now working on precise calibrations and gain tuning to optimize the performance; this should come online soon. As mentioned earlier, we started our science verification observations recently which includes detecting and characterizing companions that were analyzed in Phase I to confirm that we see an increased SNR with reduced exposure time given the Phase I improvements. These results are already looking very promising and will be published in an upcoming paper soon. We will then move towards science verification of the new KPIC observing modes such as the MDA, VFN, and PIAA before starting to use these as appropriate during regular science observation nights. We will simultaneously continue to develop the software infrastructure of KPIC to simplify the calibration procedure, user interface, and data analysis. 

Moving towards more long term tasks, there will be an upgrade to the Phase II system performed in early 2023. This upgrade will deploy the ADC as well as a laser frequency comb (LFC) meant to improve the wavelength stability during on-sky operations with wide-ranging implications on the capabilities it will enable for KPIC. The 2023 upgrade will also replace the charge 2 vortex mask currently in use for VFN with a new charge 1 mask which will double the peak coupling efficiency from 10\% to 20\% and will push to smaller angular separations; whereas the current mask has a peak coupling at $1.4\lambda/D$, the new one will peak at $0.8\lambda/D$. Further upgrades will add new fiber bundles to operate at different spectral bands, including going down to y, J, and H bands. 

\section{SUMMARY}
\label{sec:Summary}

KPIC has now been successfully overhauled from Phase I to the new Phase II system. Phase II provides several upgrades targeted at improving the stellar suppression or increasing the system throughput. These upgrades also provide new observing modes pushing for smaller angular separations. Extensive in-lab testing prior to shipping demonstrated that Phase II meets all design and science requirements, has excellent alignment, and is stable on time scales significantly longer than is required for regular operations. The system throughput and coupling with all modules on and operating matches the theoretical maximum to within measurement errors. The Phase II plate and new FEU were deployed to the Keck II Telescope in February of 2022 and commissioning began in March. Early commissioning has revalidated all elements required for Phase I operation and, as of July 2022, the Phase I capabilities have been demonstrated as fully functional on-sky. We are consistently meeting our best end-to-end throughput of 3\% from Phase I but with average conditions in Phase II. During multiple good nights in Phase II, we have demonstrated throughputs in excess of 5\% despite not using the PyWFS and high-order DM for active wavefront control which should both further improve the Strehl and hence coupling efficiency. We also successfully detected a companion at a separation of 2.4 arcsec thereby revalidating all offsetting and tracking procedures. We have started commissioning the new Phase II modes including demonstrating VFN on-sky for the first time. In the coming months, we will continue to commission the new modes and transition them to science operation. An upgrade in 2023 will further expand KPIC's capabilities and enable science at new spectral bands.


\acknowledgments 
Daniel Echeverri is supported by a NASA Future Investigators in NASA Earth and Space Science and Technology (FINESST) fellowship under award \#80NSSC19K1423. Funding for KPIC has been provided by the California Institute of Technology, the Jet Propulsion Laboratory, the Heising-Simons Foundation (grants \#2015-129, \#2017-318 and \#2019-1312), the Simons Foundation (through the Caltech Center for Comparative Planetary Evolution), and NSF under grant AST-1611623.

The authors wish to recognize and acknowledge the very significant cultural role and reverence that the summit of Maunakea has always had within the indigenous Hawaiian community. We are most fortunate to have the opportunity to conduct observations from this mountain.

\small
\bibliography{Library} 

\begin{thebibliography}{10}

\bibitem{Mawet2017_KPIC}
Mawet, D., Delorme, J.~R., Jovanovic, N., Wallace, J.~K., Bartos, R.~D.,
  Wizinowich, P.~L., Fitzgerald, M., Lilley, S., Ruane, G., Wang, J.,
  Klimovich, N., and Xin, Y., ``{A fiber injection unit for the Keck Planet
  Imager and Characterizer},'' {\em Proc. SPIE}~{\bf 10400},  1040029 (2017).

\bibitem{Mawet2017_HDCII}
Mawet, D., Ruane, G., Xuan, W., Echeverri, D., Klimovich, N., Randolph, M.,
  Fucik, J., Wallace, J.~K., Wang, J., Vasisht, G., Dekany, R., Mennesson, B.,
  Choquet, E., Delorme, J.-R., and Serabyn, E., ``{Observing Exoplanets with
  High-dispersion Coronagraphy. II. Demonstration of an Active Single-mode
  Fiber Injection Unit},'' {\em Astrophys. J.}~{\bf 838}(2),  92 (2017).

\bibitem{Bond2020_PyWFS}
{Bond}, C.~Z., {Cetre}, S., {Lilley}, S., {Wizinowich}, P., {Mawet}, D.,
  {Chun}, M., {Wetherell}, E., {Jacobson}, S., {Lockhart}, C., {Warmbier}, E.,
  {Ragland}, S., {Álvarez}, C., {Guyon}, O., {Goebel}, S., {Delorme}, J.-R.,
  {Jovanovic}, N., {Hall}, D.~N., {Wallace}, J.~K., {Taheri}, M., {Plantet},
  C., and {Chambouleyron}, V., ``{Adaptive optics with an infrared pyramid
  wavefront sensor at Keck},'' {\em Journal of Astronomical Telescopes,
  Instruments, and Systems}~{\bf 6}(3),  1 -- 21 (2020).

\bibitem{Delorme2021_KPIC}
{Delorme}, J.-R., {Jovanovic}, N., {Echeverri}, D., {Mawet}, D., {Kent
  Wallace}, J., {Bartos}, R.~D., {Cetre}, S., {Wizinowich}, P., {Ragland}, S.,
  {Lilley}, S., {Wetherell}, E., {Doppmann}, G., {Wang}, J.~J., {Morris},
  E.~C., {Ruffio}, J.-B., {Martin}, E.~C., {Fitzgerald}, M.~P., {Ruane}, G.,
  {Schofield}, T., {Suominen}, N., {Calvin}, B., {Wang}, E., {Magnone}, K.,
  {Johnson}, C., {Sohn}, J.~M., {L{\'o}pez}, R.~A., {Bond}, C.~Z., {Pezzato},
  J., {Sayson}, J.~L., {Chun}, M., and {Skemer}, A.~J., ``{Keck Planet Imager
  and Characterizer: a dedicated single-mode fiber injection unit for
  high-resolution exoplanet spectroscopy},'' {\em Journal of Astronomical
  Telescopes, Instruments, and Systems}~{\bf 7},  035006 (July 2021).

\bibitem{Morris2020_KPICPhaseI}
{Morris}, E.~C., {Wang}, J.~J., {Ruffio}, J.-B., {Delorme}, J.-R., {Pezzato},
  J., {Bond}, C.~Z., {Mawet}, D., and {Skemer}, A.~J., ``{The Keck Planet
  Imager and Characterizer: Phase I fiber injection unit early performance and
  commissioning},'' ~{\bf 11447},  1144761 (Dec. 2020).

\bibitem{Wang2021_KPICScience}
{Wang}, J.~J., {Ruffio}, J.-B., {Morris}, E., {Delorme}, J.-R., {Jovanovic},
  N., {Pezzato}, J., {Echeverri}, D., {Finnerty}, L., {Hood}, C., {Zanazzi},
  J.~J., {Bryan}, M.~L., {Bond}, C.~Z., {Cetre}, S., {Martin}, E.~C., {Mawet},
  D., {Skemer}, A., {Baker}, A., {Xuan}, J.~W., {Wallace}, J.~K., {Wang}, J.,
  {Bartos}, R., {Blake}, G.~A., {Boden}, A., {Buzard}, C., {Calvin}, B.,
  {Chun}, M., {Doppmann}, G., {Dupuy}, T.~J., {Duch{\^e}ne}, G., {Feng}, Y.~K.,
  {Fitzgerald}, M.~P., {Fortney}, J., {Freedman}, R.~S., {Knutson}, H.,
  {Konopacky}, Q., {Lilley}, S., {Liu}, M.~C., {Lopez}, R., {Lupu}, R.,
  {Marley}, M.~S., {Meshkat}, T., {Miles}, B., {Millar-Blanchaer}, M.,
  {Ragland}, S., {Roy}, A., {Ruane}, G., {Sappey}, B., {Schofield}, T.,
  {Weiss}, L., {Wetherell}, E., {Wizinowich}, P., and {Ygouf}, M., ``{Detection
  and Bulk Properties of the HR 8799 Planets with High-resolution
  Spectroscopy},'' {\em The Astronomical Journal}~{\bf 162},  148 (Oct. 2021).

\bibitem{Wang2022_KPICCORetrieval}
{Wang}, J., {Kolecki}, J.~R., {Ruffio}, J.-B., {Wang}, J.~J., {Mawet}, D.,
  {Baker}, A., {Bartos}, R., {Blake}, G.~A., {Bond}, C.~Z., {Calvin}, B.,
  {Cetre}, S., {Delorme}, J.-R., {Doppmann}, G., {Echeverri}, D., {Finnerty},
  L., {Fitzgerald}, M.~P., {Jovanovic}, N., {Liu}, M.~C., {Lopez}, R.,
  {Morris}, E., {Pai Asnodkar}, A., {Pezzato}, J., {Ragland}, S., {Roy}, A.,
  {Ruane}, G., {Sappey}, B., {Schofield}, T., {Skemer}, A., {Venenciano}, T.,
  {Kent Wallace}, J., {Wallack}, N.~L., {Wizinowich}, P., and {Xuan}, J.~W.,
  ``{Retrieving the C and O Abundances of HR 7672 AB: A Solar-type Primary Star
  with a Benchmark Brown Dwarf},'' {\em The Astronomical Journal}~{\bf 163},
  189 (Apr. 2022).

\bibitem{Sappey2022_KPICHD206893}
{Sappey}, B., {Konopacky}, Q., {Ruffio}, J.-B., {Wang}, J., {Mawet}, D.,
  {Finnerty}, L., {Schofield}, T., {Morris}, E., {Jovanovic}, N., {Skemer}, A.,
  and {KPIC Collaboration}, ``{High-Resolution Spectra of HD 206893 b with Keck
  Planet Imager and Characterizer},'' in [{\em Bulletin of the American
  Astronomical Society}{\nolinebreak\hspace{0.1em}]},   {\bf 54},  102.66 (June
  2022).

\bibitem{Wang2021_KPICPhaseI}
{Wang}, J.~J., {Delorme}, J.-R., {Ruffio}, J.-B., {Morris}, E., {Jovanovic},
  N., {Echeverri}, D., {Schofield}, T., {Pezzato}, J., {Skemer}, A., and
  {Mawet}, D., ``{High resolution spectroscopy of directly imaged exoplanets
  with KPIC},'' ~{\bf 11823},  1182302 (Sept. 2021).

\bibitem{Calvin2021_KPICLab}
{Calvin}, B., {Jovanovic}, N., {Ruane}, G., {Pezzato}, J., {Colborn}, J.,
  {Echeverri}, D., {Schofield}, T., {Porter}, M., {Wallace}, J.~K., {Delorme},
  J.-R., and {Mawet}, D., ``{Enhancing Direct Exoplanet Spectroscopy with
  Apodizing and Beam Shaping Optics},'' {\em Publications of the Astronomical
  Society of the Pacific}~{\bf 133},  024503 (Feb. 2021).

\bibitem{Ruane2018_VFN}
Ruane, G., Wang, J., Mawet, D., Jovanovic, N., Delorme, J.-R., Mennesson, B.,
  and Wallace, J.~K., ``{Efficient Spectroscopy of Exoplanets at Small Angular
  Separations with Vortex Fiber Nulling},'' {\em Astrophys. J.}~{\bf 867}(2),
  143 (2018).

\bibitem{Echeverri2019b_VFN}
{Echeverri}, D., {Ruane}, G., {Jovanovic}, N., {Hayama}, T., {Delorme}, J.-R.,
  {Pezzato}, J., {Bond}, C., {Wang}, J., {Mawet}, D., {Wallace}, J.~K., and
  {Serabyn}, E., ``{The Vortex Fiber Nulling Mode of the Keck Planet Imager and
  Characterizer (KPIC)},'' {\em Society of Photo-Optical Instrumentation
  Engineers (SPIE) Conference Series}~{\bf 11117},  111170V (Sept. 2019).

\bibitem{Guyon2003_PIAA}
{Guyon}, O., ``{Phase-induced amplitude apodization of telescope pupils for
  extrasolar terrestrial planet imaging},'' {\em Astronomy and
  Astrophysics}~{\bf 404},  379--387 (June 2003).

\bibitem{Jovanovic2017_SMFOnSky}
{Jovanovic}, N., {Schwab}, C., {Guyon}, O., {Lozi}, J., {Cvetojevic}, N.,
  {Martinache}, F., {Leon-Saval}, S., {Norris}, B., {Gross}, S., {Doughty}, D.,
  {Currie}, T., and {Takato}, N., ``{Efficient injection from large telescopes
  into single-mode fibres: Enabling the era of ultra-precision astronomy},''
  {\em Astronomy and Astrophysics}~{\bf 604},  A122 (Aug. 2017).

\bibitem{Jovanovic2020_KPICPhaseII}
{Jovanovic}, N., {Calvin}, B., {Porter}, M., {Schofield}, T., {Wang}, J.,
  {Roberts}, M., {Ruane}, G., {Wallace}, J.~K., {Bartos}, R., {Pezzato}, J.,
  {Colborn}, J., {Delorme}, J.~R., {Echeverri}, D., {Mawet}, D., {Bond}, C.~Z.,
  {Cetre}, S., {Lilley}, S., {Ragland}, S., {Wizinowich}, P., and
  {Jensen-Clem}, R., ``{Enhanced high-dispersion coronagraphy with KPIC phase
  II: design, assembly and status of sub-modules},'' in [{\em Society of
  Photo-Optical Instrumentation Engineers (SPIE) Conference
  Series}{\nolinebreak\hspace{0.1em}]},  {\em Society of Photo-Optical
  Instrumentation Engineers (SPIE) Conference Series} {\bf 11447},  114474U
  (Dec. 2020).

\bibitem{Ruane2019SPIE}
Ruane, G., Echeverri, D., Jovanovic, N., Mawet, D., Serabyn, E., Wallace,
  J.~K., Wang, J., and Batalha, N., ``{Vortex fiber nulling for exoplanet
  observations: conceptual design, theoretical performance, and initial
  scientific yield predictions},'' {\em Proc. SPIE}~{\bf 11117} (2019).

\bibitem{Wang2020_ADC}
{Wang}, J.~J., {Wallace}, J.~K., {Jovanovic}, N., {Guyon}, O., {Roberts}, M.,
  and {Mawet}, D., ``{An atmospheric dispersion corrector design with
  milliarcsecond-level precision from 1 to 4 microns for high dispersion
  coronagraphy},'' {\em Society of Photo-Optical Instrumentation Engineers
  (SPIE) Conference Series}~{\bf 11447},  1144754 (Dec. 2020).

\bibitem{NDiaye2013_ZWFS}
{N'Diaye}, M., {Dohlen}, K., {Fusco}, T., and {Paul}, B., ``{Calibration of
  quasi-static aberrations in exoplanet direct-imaging instruments with a
  Zernike phase-mask sensor},'' {\em Astronomy and Astrophysics}~{\bf 555},
  A94 (July 2013).

\bibitem{VanKooten2022_KPICZWFS}
{van Kooten}, M. A.~M., {Ragland}, S., {Jensen-Clem}, R., {Xin}, Y., {Delorme},
  J.-R., and {Kent Wallace}, J., ``{On-sky Reconstruction of Keck Primary
  Mirror Piston Offsets Using a Zernike Wavefront Sensor},'' {\em The
  Astrophysical Journal}~{\bf 932},  109 (June 2022).

\bibitem{Shaklan1988}
Shaklan, S. and Roddier, F., ``Coupling starlight into single-mode fiber
  optics,'' {\em Appl. Opt.}~{\bf 27}(11),  2334--2338 (1988).

\bibitem{Farrington2014_HD178911}
{Farrington}, C.~D., {ten Brummelaar}, T.~A., {Mason}, B.~D., {Hartkopf},
  W.~I., {Mourard}, D., {Moravveji}, E., {McAlister}, H.~A., {Turner}, N.~H.,
  {Sturmann}, L., and {Sturmann}, J., ``{Separated Fringe Packet Observations
  with the CHARA Array. II. {\ensuremath{\omega}} Andromeda, HD 178911, and
  {\ensuremath{\xi}} Cephei.},'' {\em The Astronomical Journal}~{\bf 148},  48
  (Sept. 2014).

\end{thebibliography}
\bibliographystyle{spiebib} 

\end{document}